\documentstyle[12pt]{article}
\begin{document}
\topmargin=-.5in
\oddsidemargin=.1in
\evensidemargin=.1in
\vsize=23.5cm
\hsize=16cm
\textheight=23.0cm
\textwidth=16cm
\baselineskip=24pt
\thispagestyle{empty}

\hfill{ITP-SB-98-42}\\

\smallskip

\vspace{1.0in}

\centerline{\LARGE \bf Dual Confinement Of} 

\centerline{\LARGE \bf Grand Unified Monopoles?}

\vspace{.85in}

\baselineskip=16pt{
\centerline{\large Alfred Scharff Goldhaber}
\bigskip
\centerline{\it Institute for Theoretical Physics}
\centerline{\it State University of New York}
\centerline{\it Stony Brook, NY 11794-3840}

\vspace{.5in}

\centerline{\Large Abstract}

\vspace{.5in}
\baselineskip=16pt{
\noindent\hskip.6in\vbox{\hsize 13.cm

A simple formal
computation, and a
variation on an old thought experiment, both indicate that QCD with light
quarks may confine fundamental color magnetic charges, giving an explicit as
well as elegant resolution to
the `global color' paradox, strengthening Vachaspati's SU(5) electric-magnetic
duality, opening new lines of inquiry for monopoles in cosmology, and suggesting
a  class of geometrically large QCD excitations -- loops of Z(3) color magnetic
flux
entwined with light-quark current.  The proposal may be directly testable in
lattice gauge theory or supersymmetric Yang-Mills theory.  Recent results in
deeply-inelastic electron scattering, and future experiments both there and in
high-energy collisions of nuclei, could give evidence on the existence of 
Z(3) loops.  If confirmed, they
would represent a consistent realization of the bold concept underlying the
Slansky-Goldman-Shaw `glow' model -- phenomena besides standard meson-baryon
physics manifest at long distance scales -- but without that model's isolable
fractional electric charges.} 

\baselineskip=20pt{

\newpage

\noindent
{\bf I.  Introduction}

A quarter century ago, the simultaneous and independent discoveries by Gross and
Wilczek \cite{DF} and by Politzer \cite{HDP} that quantum chromodynamics
[QCD] is asymptotically free made this theory instantly what it still is -- the
unique candidate theory for describing the structure and interactions of baryons,
as well as the mesons produced when baryons collide.  For length scales below 0.1
fm and energy scales above 1 GeV, phenomena may be described accurately by
perturbative techniques in terms of elementary quarks and gluons.  At longer
distances and lower energies, the most useful degrees of freedom become the
baryons and mesons themselves, while the connection between these two regimes is
less well determined because of the calculational difficulty associated with
nonperturbative QCD.   Nevertheless, a variety of experimental and theoretical
approaches have produced so many successes that it would seem natural to assume 
there is little or no room for surprising new phenomena in QCD.  That is
especially plausible for the perturbative regime, so if there are surprises
lurking they most likely will be found at long distance scales.  

A rare if not unique proposal in this direction, which
constitutes perhaps the boldest enterprise with Richard Slansky's name
on it, is the paper of Slansky, Goldman, and Shaw [SGS] \cite {SGS} suggesting
explicit departures from naive QCD expectations at long distance scales.  Their work
was stimulated by experimental indications of isolated electric charge with value
1/3 or 2/3 that of an electron charge.  As Gordon Shaw explains
\cite{GLS}, assuming that the manifest SU(3) gauge symmetry of QCD is reduced by a
Higgs mechanism at very long distance scales to SO(3),
one may envision isolated SO(3) singlets made of two quarks, and therefore carrying
fractional electric charge.   While seeking such objects in the laboratory
remains a worthwhile challenge, there are serious grounds to be hesitant about
the explicit SGS proposal.  The reason is that a Higgs mechanism is easy to
formulate in a regime where the gauge coupling of the theory is weak, as in the
standard model of electroweak interactions, but becomes very hard to interpret if
the coupling is strong, which inevitably would be true for QCD on the scale they
had in mind.  There was little choice about this for SGS, because any shorter
distance scale or higher energy scale with such phenomena would have been
prohibited by existing theoretical and experimental knowledge.   

Even now there is no experimental confirmation of fractional electric charge,
though searches continue \cite {GLS}, nor of any other long-range QCD effect.
This paper presents a proposal which has significant features in common with
SGS, namely, new long-range phenomena beyond ordinary hadron dynamics
(including confinement of fundamental magnetic monopoles), but does not imply
fractional electric charge.  If the proposal turns out to have merit, then it
could well be viewed as a vindication of the essence of SGS.   Certainly the
intellectual structure they developed was an important influence on my thinking.
The presentation involves both `push' and `pull' heuristic arguments, i.e., 
reasons
to suspect the existence of new phenomena as well as appealing consequences
which would follow if they occurred, but does not include a proof that they are
inevitable.  Because of the wide range of application, there likely will be a number
of ways to test the proposal, including several outlined later.  The focus
begins with particles dual to quarks, namely, fundamental magnetic monopoles
carrying both ordinary and color magnetic charge.

Before plunging into the proposal, we should review some long-range effects
which already are expected, resulting from hybridization between different scales.
If a region of space is sufficiently hot, then the temperature $T$ sets a scale
which invokes asymptotic freedom, and thus allows one to describe the properties
as if dealing with a gas of free quarks and gluons, commonly known as the
`quark-gluon plasma'.  Clearly long-distance correlations in this regime should
look quite different from those at $T=0$.  Thus, long-distance phenomena are
altered in a predictable way, but in a high-energy rather than low-energy
regime.  Something similar should happen if, for example, compression of
cold nuclear matter, as in the interior of a neutron star, were to produce baryon
density much greater than that in normal nuclei.  This time the high density gives
a short-distance and therefore high-energy scale which implies asymptotic freedom
and a change in long-distance correlations.  

Neither of these effects would be
a surprise. Something a bit closer was the proposal of Jaffe \cite {RJ} that a
six-quark system of strangeness $S = 2$ might be stable against decay to two
$\Lambda$ particles.  This in turn raises a possibility discussed by Witten \cite
{jew} that electrically neutral strange matter might be stable or metastable. 
However, even this effect if it occurred would be a consequence of relatively
short-range interactions.  Something still closer to what follows is the
suggestion \cite{cgjj} that nuclei might have stable or metastable toroidal
forms, where a quark-containing tube of color-magnetic flux is bent into a closed
curve, a structure which might exhibit rigidity and incompressibility as well as
tension.
\bigskip
\eject
\noindent
{\bf II.  Heuristic arguments for monopole confinement}

\noindent
{\it Gauge invariance}

Though the existence of a magnetic monopole has yet to be confirmed by
experimental observation, even as a concept this object repeatedly has
played the role of an intellectual {\it aqua regia}, exposing profound
aspects of structure in physical systems.  The most noted example is
Dirac's realization \cite {Dirac} that the existence of isolable monopoles would
require the quantization of electric charge $q$ and magnetic charge $g$, through
the quantum condition that the product of $q$ with $g$ is proportional to an
integer.   

Dirac monopoles `inserted' into QED give a model for confinement, because emanating
from an elementary pole inside a Type II superconductor with its electron-pair
condensate 
must be two strings of superconductor-quantized magnetic flux,
each terminating only on an antipole.  If one pair of pole and antipole were slowly
separated, clearly the two strings coming out of the pole would terminate on the 
antipole, implying a confining string tension holding the two together.    
By analogy, if the vacuum of QCD without
light  quarks comprised a monopole condensate  \cite {mand, 't H}, this
then would confine heavy quarks.  However, if there are elementary quarks light on
the scale of $\Lambda _{QCD}$, then there
is no confinement of heavy quarks:  Instead pair creation of light quarks
allows heavy-quark-containing mesons to be separated with no further energy
cost.  

A general argument of 't Hooft \cite { 't H2}  for QCD without light quarks
shows that either heavy elementary quarks or heavy fundamental monopoles
should be confined, but not both.  If this argument still applied in the
presence of light quarks, then monopole  confinement would be a triviality. 
In any case this makes it clear that there would be nothing obviously
inconsistent about such confinement.  To understand why it might
be expected,
let us examine in the more familiar superconductor case the issue of
screening, and what charges can or cannot be screened.   In electrodynamics
it is useful to distinguish two different kinds of
conserved charge, local or Gauss-law charge and Aharonov-Bohm
[AB] or Lorentz-force charge  \cite {GK}.  Although the
local charge of an electron-quasiparticle is completely screened inside a
superconductor, the AB charge cannot be screened \cite {RA},
because of the reciprocity requirement that an AB phase of $\pi$ 
must occur whether a quasiparticle is diffracted around a fluxon (i.e., a
superconductor quantum of flux) or a fluxon
is diffracted around a quasiparticle.    If the effect on the fluxon is to be
described by a local interaction, evidently the AB charge is not screened.  

Now let us look at the same issue for a fundamental monopole in QCD.  
The monopole may be characterized as the source of a Dirac string 
carrying color magnetic flux which would produce an AB phase $2\pi /3$ for a
fundamental quark diffracted around it  Of course, in addition to 
this color flux there must be an ordinary
magnetic flux in the string yielding a phase $4\pi/3 \ {\rm mod}(2\pi)$.   The
fractional color flux in the string implies that there must be a net nonzero color
magnetic flux coming out  of the pole.  A monopole whose Dirac string would carry
full
$2\pi $ color flux has no such consequence, because that could be exactly
compensated by an `adjoint'
 monopole made from a classical configuration of purely 
SU(3) gauge fields.  Thus, in the sense just
defined, adjoint monopoles can be screened but fundamental monopoles cannot
(a suggestive analogue to what happens with adjoint gluons and
fundamental quarks).

  In the absence of dynamical quarks,
this lack of screening may not matter, because vacuum fluctuations in the form of 
loops carrying flux
$2\pi /3$ occur easily on arbitrarily large length scales, so that the magnetic
charge is not definable as an eigenvalue.  In this sense, it may be screened 
just like electric charge in a normal metal, i.e., with mean value zero but such
large fluctuations that it is not defined as a sharp quantum observable.  Thus the
monopole charge, rather than being screened or compensated, may be hidden in much
the same way a needle becomes invisible inside a haystack.  Here is another
perspective:  In the theory with only adjoint fields, such as those of the gluons,
the gauge symmetry is SU(3)/Z(3), so that an arbitrarily thin tube of Z(3)
magnetic flux would be invisible even by the AB effect
to all
elementary excitations, hence could not excite the vacuum, and therefore need
not carry an observable energy per unit length, as would have to be true for an
observable string.

On the other hand, once quarks are present, there could be a
nonvanishing string tension for loops of Z(3) color magnetic flux, so that
geometrically large quantum fluctuations of these loops should be suppressed.
A reason for suspecting this is that now the AB effect would make even the thinnest
tube visible for those quark trajectories which surround the tube. Thus, the
fractional color magnetic charge could become sharp, meaning that an observable
color magnetic field, confined to a tube of fixed radius, 
emanates from the monopole out to infinity. More formally, because now there are
particles in the fundamental representation of SU(3), the full gauge symmetry
applies, and so a nonzero Z(3) color magnetic flux out of a fundamental monopole is
at least potentially observable.   

\noindent
{\it Ideal experiment}

Here is a thought experiment suggesting the same conclusion.  Imagine a hadron such
as a proton at rest near an SU(5) monopole, with its ordinary as well as color
magnetic charge.  If a deeply-inelastic electron scattering sends a quark out
of the proton with very high momentum parallel to that of the incident electron,
then the quark's evolution in the beam direction can be described perturbatively  for
a time proportional to that momentum, which means the ordinary magnetic field of the
monopole will deflect it in such a way that only a fraction of a quantum of angular
momentum will be transferred to the quark.  This is inconsistent with
conservation and quantization of
angular momentum \cite{G}.  The analysis also can be carried out in the rest
frame of the fastest final hadron.  In this frame a perturbative computation is
accurate for a fixed time of order $1 \ {\rm fm/}c$.  However, the magnetic field
of the monopole flashes by the quark in a much shorter time because of the 
Lorentz contraction of the field configuration, and therefore again there is a
definite, but fractional transfer to the quark of angular momentum projected along
the beam direction.     

 One way to restore consistency is to assume that there is also a
spherically symmetric color magnetic field, so that the combined fields
always transfer an integer number of angular momentum units  to the quark. 
However, that assumption directly contradicts the most basic understanding of
QCD, which requires a mass gap for color-carrying excitations, so
that a long-range, `classsical', isotropic,
color-magnetic field is impossible.
How can these two requirements, of nonscreening and yet no isotropic
long-range field, be reconciled?  An obvious if not unique way to
avoid the dilemma is by `escape into asymptotic freedom':  Color fields make
sense in the high-energy, short-distance, perturbative regime, so if the magnetic
flux comes out in a tube with radius of scale  $ \le \Lambda _{QCD}^{-1}$ it is
consistent with knowledge about the low-energy behavior of the theory, and at the
same time satisfies the requirement of nonscreening.   Evidently such a tube must
have a finite tension, so that the energy of a pole-antipole pair connected
by the tube
must rise linearly with separation, and this implies confinement of
fundamental monopoles.

From the viewpoint of the deeply-inelastic `thought experiment', why
shouldn't the monopole confinement argument apply even if there
are no dynamical quarks? In that case heavy `external' quark sources certainly are
confined, and the failure of angular momentum quantization for a single
pole-quark pair is acceptable, as there is never a single isolated quark moving in
the field of the monopole.

While the above arguments might be appealing, they surely do not constitute
a proof of monopole confinement.  The reason is that even with light quarks
it may be that large loops of magnetic flux, at least of a certain cross
sectional radius, still have arbitrarily low energy, in which case they would be part
of the vacuum structure rather than physical excitations.  Then the net flux out
of a monopole again would be hidden by vacuum fluctuations.  However, because the
quarks would be sensitive to arbitrarily thin tubes even with Z(3) flux, there is
now a much stronger constraint, from below as well as above, on the acceptable
radii for flux tubes with very low energy.   

Both because the thought experiment was the germinating element in my own thinking
on this subject, and because more careful examination could tend either to strengthen
or to weaken the argument, it seems worthwhile to focus more explicitly on the wave
function evolution entailed by this process.  In the presence of gauge fields, the
conventional (non-gauge-invariant) momentum of an object whose charges couple to
these fields becomes undefined.  Thus, in the directions transverse to the very high
momentum of the struck quark, it makes no sense to think about the momentum of the
quark by itself. However, the correlated wavefunction of the quark and the
associated slower remnants might have a well-defined wave function in transverse
momentum $p_T$, a wave function which initially would be strongly peaked at
$p_T\approx 0$  and azimuthal angular momentum about the beam direction also zero. 
Then the essential idea is that, absent any contribution from color magnetic
fields, the only effect feeding some change in this azimuthal angular momentum
would be coming from the scattering of the fast quark on the ordinary magnetic field
of the monopole.  A fractional value for this angular momentum transfer gives the
conclusion that something is inconsistent about this picture, and leads by
elimination of alternatives to the
inference that an observable string of color magnetic flux emanates from the
monopole.   

If we accept that inference, how do we find consistency restored?  As an example,
imagine that the observable string comes out of the monopole in the direction
parallel to the fast quark momentum.  We still may use gauge invariance to
place the Dirac string of
ordinary plus color flux anywhere we like, and thus may choose it along the
observable color string.  In this case, for all except those trajectories which
penetrate the observable, finite-thickness string, the effective field is just that
of a pole which is one end of a solenoid with ordinary magnetic flux such that a
quark going around it acquires a fractional phase $2\pi/3 \ {\rm mod} \ 2\pi$. 
Evidently mesons or baryons generated by fragmentation of the fast quark will always
have integer azimuthal angular momentum, but nevertheless the initial effect of fast
passage of the quark by the monopole will be to generate a fractional change in
the net angular momentum of the entire system interacting with the pole,
something now allowed because an observable string with fractional magnetic flux
is present.      

\bigskip

\noindent
{\bf III.  Consequences and applications}

Now let us look at how  fundamental monopole confinement
would reorient perspectives on a variety of issues.  

\noindent
1. {\it Paradox of `global
color'}

  A number of authors addressed the problem of generalizing a
collective-coordinate quantization technique, accepted as describing the electric
or `dyon' charge of an SU(2) monopole, to the case of the
SU(5) monopole  \cite{Ab,Bal,NM,CN}.  They found that for the unbroken SU(3) of color the
dyon charge of an isolated pole is not defined -- an effect reminiscent of
spontaneous symmetry breaking as in ferromagnetism.  Evidently if monopoles with
fundamental color charge are confined, this problem simply disappears. 
A more general and straightforward comment is that, with or without
monopole confinement,
 the paradox  is ill-posed, because
the collective-coordinate method has been used to quantize zero modes of the
monopole placed in a perturbative QCD vacuum, which definitely is an
incorrect description of the lowest-energy degrees of freedom on length scales large
compared to $\Lambda _{QCD}^{-1}$.   Thus, while monopole confinement eliminates
the problem at the very beginning, the significance of that resolution perhaps
is diminished
because there might well not be such a problem if the right vacuum were
understood well enough to be implemented for the analysis.

\noindent
2.  {\it Electric-magnetic duality in a grand unified model}

Recently Vachaspati \cite{V} has described a remarkable duality of
SU(5), clearly relevant for any grand unified theory.  The fundamental
monopole is part of a family of tightly bound states, with
magnetic charges 1,2,3,4, and 6 times the fundamental charge.  These five
states can be identified as dual partners of the five
fundamental fermions in SU(5), three quarks, a lepton, and a neutrino. 
There is a possibly deep or possibly just technical issue, that the charge-2
state should be identified with an antiquark.  There are two other
difficulties.  First, the monopoles appear to be
spinless, while the fermions of course have spin-1/2.  This problem arose
already with the original Montonen-Olive proposal of duality between
monopoles and gauge bosons \cite{MO}, and eventually 
found two resolutions.  One is to introduce supersymmetry, so that both
monopoles and the dual elementary particles come in families with the same
range of spins \cite{WO,HO,sen,SW1,SW2}.  The other, acknowledging the possibility
in principle of making a perfect correspondence through supersymmetry, is to
be satisfied with what might be called `virtual duality' -- a symmetry
applied to all properties except spin.  Whichever approach one prefers,
with respect to this issue
Vachaspati's system is in the same category as the older examples.

The other difficulty \cite {V} is that the effective long-range couplings of the
monopoles and their dual partners are identical, except that quarks are
confined, whereas previous discussions suggested that the colored monopoles
are not.   For this reason Vachaspati considered introducing the confinement
essentially by hand.  The argument above that the monopoles which nominally carry
nontrivial
Z(3) color magnetic charge are automatically confined gives a
way to
perfect Vachaspati's duality, lending additional interest to pursuing it
further.  One side note:  Confinement could lead to loosely bound
`baryons', but these then could collapse to the tightly bound `leptons'
already identified in Vachaspati's scheme.  Clearly this is different from
the separate baryon and lepton conservation laws which apply at low energy
scales, but as one expects those laws not to hold for particles on energy scales
approaching the monopole masses this may well be a consistent result. 

\noindent
{\it Monopole evolution in cosmology}

  Monopoles formed on a
mass scale  significantly higher than the mass scale for inflation would have
disappeared during inflation \cite {Guth}; indeed, that is one of the attractive
features of inflationary models.  However, lighter monopoles would need some other
mechanism to explain why we don't see abundant evidence of their existence today.  
Many such mechanisms have been proposed, up to quite recent times.  One possibility
is that the dynamics at some intermediate era between monopole formation and the
present would make the poles  unstable, allowing them to disappear, even though any
remnant which did survive would be stable now
\cite{LP,D}.   

If monopoles were created at some early epoch and not swept away
meanwhile, then the only way to explain their scarcity today would be by
confinement, exactly the phenomenon discussed here.  How would that work?  If
monopoles were formed above the QCD phase transition expected at a temperature of
order $\Lambda_{QCD}$, then confinement below that transition would result in
attachment of Z(3) strings to each pole, either a single outgoing $2\pi /3$
string, or two outgoing $-2 \pi /3$ strings, with the opposite arrangement for
antipoles.  A pair connected by a single string likely would have disappeared by
now, thanks to dissipative forces leading to gradual collapse
and annihilation.  On the other hand,
a large loop with alternating negative and positive flux connecting alternating 
pole and antipole could be much more durable.  This kind of `cosmic necklace', with
the poles as `beads', was suggested by Berezinskii and Vilenkin \cite {BV} as a
possible source of the highest-energy component of the cosmic ray spectrum, through
occasional annihilations of poles and antipoles, which might for example
slowly drift together by sliding along the string.  The evolution of networks of
such strings is an interesting and nontrivial problem, which could be studied once
the basic couplings associated with string crossings were determined.  In
particular, in principle a `fusion' of three strings converging together should be
possible, which would allow three monopoles to be connected to each other in a
dual version of a baryon.   However, if this could happen easily then the problem
of too many monopoles would be
restored, so a crucial question is whether there is a substantial
inhibition of such fusion.  

\bigskip

\noindent
{\bf IV.  New phenomena in QCD at accessible scales}

\noindent
{\it Theoretical aspects}

Even though it is consideration of heavy, fundamental monopoles and their
interactions which has led here to the suggestion that they would be held together by
Z(3) color flux strings,  that statement clearly 
has a consequence for phenomena at much lower scales
than the monopole mass.  It means that even in the absence of such poles QCD must
support excitations consisting of loops of color magnetic flux, with the mass of 
a loop being proportional to its circumference.   The loops would be unstable
against shrinkage, but would give an interesting and nontrivial structure of QCD
excitations on a length scale large compared to $1/\Lambda_{QCD}$.  This is
reminiscent of the Slansky-Goldman-Shaw proposal to explain
experimental reports of fractional electric charge \cite{SGS}.  As mentioned
earlier, they noted that if a Higgs mechanism at energy scales below, or length
scales above, the scale associated with $\Lambda_{QCD}$ could operate to reduce
SU(3) of color to SO(3) of `glow', then diquarks could exist in isolation, and
of course would carry fractional electric charge.  Shortly after, Lazarides, Shafi,
and Trower [LST]
\cite{LST} observed that such a Higgs effect automatically would imply confinement
of fundamental monopoles exactly like that argued above.   

As was also mentioned earlier, there is no natural starting
point from which the phenomena of this particular Higgs mechanism could be deduced
in a perturbative framework, the only recognized way to do it.  This criticism
applies equally to the deduction by LST.  Of course, the fact that a conceivable
route to a particular result turns out to be rocky and uncertain does not
mean the result itself is necessarily wrong, only one still lacks evidence that it
is right.    Here the issue has been approached from the other end, and fundamental
monopole confinement derived. This does not necessarily imply the isolability of
fractional electric charge or the screening of some QCD color-electric fields, but
it certainly does say there must be a new feature of QCD at large length scales,
namely, loops of color-magnetic flux, just as indicated by LST.  Without light
quarks, heavy quark confinement implies loops of color-electric flux, so familiar
pictures would not be changed so enormously, just `dualized'.  This means that the
change in structure of QCD as the mass of light quarks passes from above to below
$\Lambda_{QCD}$ would be quite subtle:  Above there would be at least metastable
color-electric but substantial suppression of color-magnetic strings (more
accurately, very low magnetic string tension), and below something more like the
opposite would be true.

The meaning of confinement or non-confinement needs a bit more attention.  In terms
of a four-dimensional euclidean path integral, confinement is associated with
exponential suppression, with the area of an appropriate loop appearing in the
exponent, as opposed to effects associated with widely
separated finite-mass excitations, in which case only the length (perimeter) of the
loop appears.  For QCD with dynamical quarks, sufficiently large loops
must exhibit a perimeter law, but the coefficient of the perimeter term itself falls
exponentially with quark mass because the tunneling leading to quark pair creation
is exponentially suppressed.  Thus a visible transition on a finite lattice from
area to perimeter law occurs at some finite mass, presumably of order
$\Lambda_{QCD}$, and should be rather smooth.   For the proposed monopole
confinement, with monopoles expected to be extraordinarily massive, the breaking of
strings by monopole pair creation should be impossibly rare for observation on any
finite lattice.  If the magnetic strings only exist for finite quark mass, it
becomes a delicate question exactly how the string tension depends on that mass. 
However, again one would expect a smooth transition, with the maximum tension
approached for quark mass below 
$\Lambda_{QCD}$.  This leads to the amusing conclusion that fundamental dyons
carrying both monopole and quark charges might exhibit an effective confinement 
with very weak dependence on quark mass.

 If all this were confirmed, it would be a vindication of the essential
claim of SGS for nontrivial manifestations of fundamental
QCD degrees of freedom at large length scales.
These color-magnetic-flux-loop excitations presumably should be an important
class of what have been called `glueballs', which likely would be drastically
different in character from what one would find in QCD without light quarks, and
would NOT be pure glue, as the light quarks must be an essential part of their
structure.  

As already stated, the fact that confinement of fundamental  monopoles would be
equivalent to the existence of
Z(3) magnetic flux strings means that there is a way to test this proposal in
familiar energy regimes of QCD.  In particular, as lattice calculations grow 
steadily better at taking
account of light quark degrees of freedom, it should become possible to study this
issue on the lattice and obtain credible results.  The best way to
formulate the problem might be to insist $\grave{\rm a}$ la Wu and Yang \cite{WY}
that along a straight line between monopole and antimonopole there is a
gauge-matching
between vector potentials outside and inside the smallest
plaquettes surrounding that line, involving one unit of Z(3) color flux, and
one Aharonov-Bohm unit of ordinary magnetic flux.  This means a 
phase of $2\pi$ associated with those plaquettes for $u$ quarks encircling them,
but $0$ for $d$ quarks.  Of course in all other respects there is a standard Dirac
electromagnetic monopole vector potential for the pole-antipole system. All this
implies, as stated earlier, that there must be a net Z(3) color magnetic flux
between monopole and antimonopole.  If that flux were observable and not hidden,
then monopole confinement would follow, and would be signaled by an  
area law for exponential suppression of monopole loops in
the path integral, associated with the product of the separation between pole and
antipole and the Euclidean time duration of that separation.

There might be an analytic approach to determining whether or not confinement
occurs, afforded by recent progress in studying supersymmetric nonabelian gauge
theories \cite{SW1,SW2,IS,P}.  In these theories it is often possible to make
precise conjectures about the particle spectrum, and to verify the conjectures not
by a direct proof but rather by subjecting the proposed forms to many different 
consistency checks, and finding that all are passed.  To do this for our problem
would require starting with at least an SU(5) theory (including a
hypermultiplet containing quarks and leptons), and following an elaborate sequence
of Higgs mechanisms to break the manifest symmetry down to ${\rm SU(3)_{color}
\times U(1)_{electromagnetic}}$.  This is surely much more complicated than
anything which has been done so far with such systems, but might nevertheless be
manageable. 

\noindent
{\it Experimental aspects}

As physics is an experimental science, it surely is worth considering how the new
kind of structure proposed here might be accessible to experimental observation.
Up to this point in the paper, the main speculation has been the unproved
proposal that Z(3) flux loops may exist.  To connect that with experiment entails
more speculation.  Conventional hadron collisions are not promising.   First of
all, any frequently occurring peculiar phenomena in such processes would have been
noted already.  Secondly, because Z(3) strings cannot break by creation of
light-quark pairs, their coupling to conventional hadrons should be weak.  This
implies that they would not be generated easily in typical collisions.  What
couplings would be possible?  Because $u$ and $d$ quark vacuum currents would
circulate oppositely around the string, there should be a $\rho_0$-meson magnetic
coupling `contact term' -- i.e., only acting on sources which themselves overlap
geometrically with the string.  Thus a contracting string could release energy
through emission of $\rho_0$ mesons, but as these are rather massive there would
be a poor match, between the likely small energy release in the contraction from
one loop energy eigenstate to a lower one and the large mass of the emitted
particle.  All this implies a quite substantial lifetime for a large loop before
collapse.  Slow decay almost invariably goes together with low production rates,
and helps to explain why even if they can exist Z(3) loops would not have leapt
to our attention.

Recently, experiments on deeply-inelastic electron-proton
scattering \cite{HZ} have been interpreted as indicating a `hard-pomeron'
contribution to the reaction \cite{DL,H,ZZ}.  By familiar reasoning of Regge
duality, such an effect should be associated with a new class of glueball
excitations
\cite{DL}.  Could these new glueballs be the magnetic loops proposed here?  If so,
then it would not be strange if processes described by the hard pomeron also
produced the loops. Perhaps  detailed exclusive or semi-inclusive studies of
such events would reveal structure related to the loops, formed
as geometrically large and therefore high-energy excitations.  

It then becomes interesting to consider what
kind of signal such an object would generate, but that is not easy to determine.
All features of closed-string dynamics, many still obscure despite all the years
of string studies, would appear to be relevant for the behavior of these Z(3)
loops.  Thus some caution is needed in guessing what should happen in these
scattering processes.   This time of course the coupling leading to production
would be electromagnetic, but again would involve a contact interaction which at
lowest order in momentum transfer would be to the anapole or toroidal moment of the
flux loop.  This suggests that at the moment of appearance the loop would be quite
small in size, but then could expand.  If
such primitive thinking covers the main features, then it becomes possible to
suppose that in a suitable frame boosted along the beam direction there would be a
fairly large isotropic ensemble of pions. Because of the decay energy mismatch
mentioned earlier, the pions might be quite limited in their range of momenta. 
Such an effect could be quite striking, and very different from typical results of
deeply-inelastic scattering.  

A different picture, analogous to bremsstrahlung of
photons, would be that with small probability virtual Z(3) loops exist in the
neighborhood of the incident proton, and these are made real by the absorption of
the highly virtual photon.  To avoid enormous form-factor suppression, in a
suitable Breit frame the initial and final momenta of the loop would both have to
be large, implying Lorentz contraction which compensates for spatial oscillation of
the phase factor in position space, an effect discussed some time ago for elastic
scattering on deuterons
\cite {CW}. 

If deeply-inelastic electron-proton scattering gives an indirect hint
of new long-distance dynamics in QCD, plus the potential to provide more direct
evidence, then very high-energy nucleus-nucleus collisions at least have
the possibility of generating such evidence in processes with large rate.  By
heating substantial volumes above the QCD phase-transition temperature, such
collisions could permit formation of Z(3) loops, and  if so their 
slow decay
during the cooling process would give a characteristic signal, providing
important evidence that quark-gluon plasma had formed. If the 
idea of slow decay is right, this 
would allow a loop to escape from the dense, highly excited formation region,
and then to populate a small volume in pion momentum space with a large
number of particles. 

While these thoughts about experimental signals are vague and sketchy, it may
well be possible with further study to make more precise statements.  What seems
likely to be unchanged is the fact that a geometrically large object of high
coherence, which decays slowly in small energy steps, will produce a signal
different from any more familiar system, including a large nucleus.  [Of course, if
the objects turned out to be at least stable against small perturbations, the
signal would be even more striking.]

\bigskip

\noindent
{\bf V.  Conclusions}

In the title the phenomenon proposed here is referred to as `dual confinement'.
Of course this makes sense because familiar color-electric confinement of heavy
quarks is replaced by color-magnetic confinement of heavy monopoles, but if the
proposal is correct as stated then something deeper is at work.  Usual discussions
posit a duality between superconducting screening of one kind of charge and
confinement of the other. Here, however, the screening of color-electric charge is
more powerful even than that by a superconductor, because for every heavy quark
there is an attached light antiquark, exactly screening not only the local charge
but also the Aharonov-Bohm charge.   Total screening of the color electric charge
carried by heavy quarks is the remnant of the heavy-quark confinement which exists
without light quarks. Thus the duality would be one between confinement of
fundamental color-magnetic monopoles and total screening of heavy quark
color charge, not inconsistent with the familiar version but nevertheless clearly
different.  If found, such a duality therefore would be something new.   

It is
enticing to think that physics research is now at a stage where within a short
time  there might be direct evidence from a variety of directions on whether Z(3)
strings occur in nature.  Lattice gauge theory or supersymmetric gauge theory
could give information, as could deeply-inelastic electron scattering or
high-energy nucleus-nucleus scattering.  A positive answer would provide a firm
foundation for the theoretical and cosmological applications explored above. 
Perhaps even more satisfying if this happened would be the realization of a
remarkable new consequence of QCD.  This suggests a further challenge:  Are there
any other possible ways in which QCD could really give us a surprise?  Not easy or
obvious, but surely worth looking!

This study was supported in part by
the National Science Foundation.  I have benefited over a period of time
from conversations with Martin Bucher, Georgi Dvali, Edward Shuryak, Mikhail
Stephanov, and Tanmay Vachaspati.  Richard Slansky was a valued friend and
colleague from student days on -- about forty years.  Although we never
collaborated on a paper, it was always a pleasure to discuss with him, and to
experience his intelligent and discriminating enthusiasm for physics.  His
courage in facing physical challenges (in all senses!) was inspiring.  Truly the
word ``glow'' was as descriptive of his luminous personality as of the beautiful
SGS idea.

\end{document}